\documentstyle[psfig,epsf]{article}
\begin{document}
%
\def\lta{\;\raisebox{-.5ex}{\rlap{$\sim$}} \raisebox{.5ex}{$<$}\;}
\def\gta{\;\raisebox{-.5ex}{\rlap{$\sim$}} \raisebox{.5ex}{$>$}\;}

%

\newcommand{\permille}{$^0 \!\!\!\: / \! _{00}\;$}
\newcommand{\GeV}{GeV}
 
\newcommand{\mt}{m_{t}}
\newcommand{\mtt}{m_{t}^2}
\newcommand{\mw}{M_{W}}
\newcommand{\mww}{M_{W}^{2}}

\newcommand{\md}{m_{d}}
\newcommand{\ms}{m_{s}}
\newcommand{\mb}{m_{b}}
\newcommand{\mbb}{m_{b}^2}
\newcommand{\mc}{m_{c}}
\newcommand{\mh}{m_{H}}
\newcommand{\mhh}{m_{H}^2}
\newcommand{\mz}{M_{Z}}
\newcommand{\mzz}{M_{Z}^{2}}

\newcommand{\lra}{\leftrightarrow}
 
\newcommand{\ie}{{\em i.e.}}
\def\Ww{{\mbox{\boldmath $W$}}}  
\def\B{{\mbox{\boldmath $B$}}}         
\def\nn{\noindent}

\newcommand{\sinsq}{\sin^2\theta}
\newcommand{\cossq}{\cos^2\theta}
\newcommand{\be}{\begin{equation}}
\newcommand{\ee}{\end{equation}}
\newcommand{\ba}{\begin{eqnarray}}
\newcommand{\ea}{\end{eqnarray}}

\newcommand{\nl}{\nonumber \\}
\newcommand{\eqn}[1]{Eq.({#1})}
\newcommand{\ibidem}{{\it ibidem\/},}
\newcommand{\into}{\;\;\to\;\;}
\newcommand{\wws}[2]{\langle #1 #2\rangle^{\star}}
\newcommand{\p}[1]{{\scriptstyle{\,(#1)}}}
\newcommand{\ru}[1]{\raisebox{-.2ex}{#1}}
\newcommand{\epem}{$e^{+} e^{-}\;$}
\newcommand{\tcht}{$t\to c H\;$}
\newcommand{\tczt}{$t\to c Z\;$}
\newcommand{\tcgt}{$t\to c g\;$}
\newcommand{\tcft}{$t\to c \gamma\;$}
\newcommand{\tchm}{$t\to c H$}
\newcommand{\tczm}{$t\to c Z$}
\newcommand{\tcgm}{$t\to c g$}
\newcommand{\tcfm}{$t\to c \gamma$}
\newcommand{\tch}{t\to c H}
\newcommand{\tcz}{t\to c Z}
\newcommand{\tcg}{t\to c g}
\newcommand{\tcf}{t\to c \gamma}

\newcommand{\tbwh}{t\to b W H}
\newcommand{\tbwz}{t\to b W Z}
\newcommand{\tbwht}{$t\to b W H\;$}
\newcommand{\tbwzt}{$t\to b W Z\;$}
\newcommand{\tbwhm}{$t\to b W H$}
\newcommand{\tbwzm}{$t\to b W Z$}

\newcommand{\tbwf}{t\to b W \gamma}
\newcommand{\tbwg}{t\to b W g}
\newcommand{\tbwft}{$t\to b W \gamma\;$}
\newcommand{\tbwgt}{$t\to b W g\;$}
\newcommand{\tbwfm}{$t\to b W \gamma$}
\newcommand{\tbwgm}{$t\to b W g$}

\newcommand{\tcww}{t\to c W W}
\newcommand{\tcwwt}{$t\to c W W\;$}
\newcommand{\tcwwm}{$t\to c W W$}

\newcommand{\tuww}{t\to u W W}
\newcommand{\tuwwt}{$t\to u W W\;$}
\newcommand{\tuwwm}{$t\to u W W$}

\newcommand{\tqw}{t\to q W}

\newcommand{\tbw}{t\to b W}
\newcommand{\tbwt}{$t\to b W\;$}
\newcommand{\tbwm}{$t\to b W$}

\newcommand{\tsw}{t\to s W}
\newcommand{\tswt}{$t\to s W\;$}
\newcommand{\tswm}{$t\to s W$}
\newcommand{\tdw}{t\to d W}
\newcommand{\tdwt}{$t\to d W\;$}
\newcommand{\tdwm}{$t\to d W$}

\newcommand{\Gt}{\Gamma(t\to b W)}

\newcommand{\ttbar}{\mbox{$t \bar{t}$}}
\newcommand{\lolumi}{\mbox{$\rm {10^{33}~cm^{-2}\,s^{-1}}$}}
\newcommand{\hilumi}{\mbox{$\rm {10^{34}~cm^{-2}\,s^{-1}}$}}
\newcommand{\pT}{\mbox{$p_{T}$}}
\newcommand{\invfb}{\mbox{fb$^{-1}$}}
\newcommand{\ra}{\mbox{$\rightarrow$}}
\newcommand{\Zll}{\mbox{$Z \rightarrow l^{+}l^{-}$}}
\newcommand{\mll}{\mbox{$m_{l^{+}l^{-}}$}}
\newcommand{\mZ}{\mbox{$m_{Z}$}}
\newcommand{\ET}{\mbox{$E_{T}$}}
\newcommand{\lplm}{\mbox{$l^{+}l^{-}$}}
\newcommand{\modeta}{\mbox{$\mid \eta \mid$}}
\newcommand{\Wlnu}{\mbox{$W \rightarrow l\nu$}}
\newcommand{\pTmiss}{\mbox{$p_{T}^{miss}$}}
\newcommand{\mqZ}{\mbox{$m_{qZ}$}}
\newcommand{\Ldt}{\mbox{$\int{\cal L} \cdot {\rm dt}$}}
\newcommand{\Wjj}{\mbox{$W \rightarrow jj$}}
\newcommand{\Wpl}{\mbox{$W^{+}$}}
\newcommand{\csq}{\mbox{$c^{2}$}}
\newcommand{\ETmiss}{\mbox{$E_{T}^{miss}$}}
 
\begin{center}
{\Large \bf Top quark rare decays  
in the standard model and beyond} \\

\vspace{4mm}

Barbara Mele\\
INFN, Sezione di Roma and University of Rome "La Sapienza", Rome,\\
 Italy\\
\end{center}

\begin{abstract}
\noindent
Theoretical predictions for the top quark rare decays are reviewed 
within and beyond the standard model.
Expectations at the CERN Large Hadron Collider are discussed.
\end{abstract}


\section{Introduction}  

The production of $10^7-10^8$ top quark pairs per year at 
the CERN Large Hadron Collider (LHC) will
allow to probe the top couplings to both  known and new particles
involved in possible top decay channels different from the main
$\tbw$. Thanks to the large top mass, there are several decays 
that can be considered, even involving the presence of on-shell heavy 
vector bosons or heavy new particles in the final states.
On a purely statistical basis, one should be able to detect
a particular decay channel whenever its branching ratio (B) is larger than
about $10^{-6}-10^{-7}$. In practice,  
background problems and systematics will
lower this potential by a few orders of magnitude, the precise reduction
being dependent of course from the particular signature considered.
We will see, that the final detection threshold for each channel will  
not allow  the study of many possible final states predicted in the 
standard model (SM), unless new stronger couplings come into play.

After discussing in section 2 the main top quark rare decay channels 
predicted by the SM, we review in section 3 the  
expectations of a large class of possible extensions of the SM, such as
the inclusion of a 4$^{th}$ fermion family, the two-Higgs-doublet models,
and the minimal supersymmetric standard model (MSSM).
Finally, results of the analysis of the LHC potential in the 
top rare decays field are reported in section 4.

\section{Standard Model Decays}
\label{bms:sm}
In this section, we give an overview of the decay channels of the 
top quark in the framework of the SM.
In the SM the decay $t \to bW$ is by far the dominant one. The top total
width is then given by:
\be
\Gamma_T \simeq \Gamma(t \to bW) 
\simeq 0.17 \,\,{\rm GeV} |V_{tb}|^2 \frac{m_t^3}{M_W^3} 
\simeq 1.55 {\rm GeV}.
 \nonumber
\ee
The rates for other decay channels are predicted to be smaller  by
a few orders of magnitude in the SM.  
The second most-likely decays are the Cabibbo-Kobajashi-Maskawa
(CKM) non-diagonal decays \tswm and \tdwm.
Assuming $|V_{ts}| \simeq 0.04$ and $|V_{td}| \simeq 0.01$, 
respectively~\cite{Caso:1998tx}, 
one gets;
\be
B(\tsw) \sim 1.6 \times  10^{-3} \; \;{\rm and} \; \;
B(\tdw) \sim 1 \times 10^{-4},
\label{bme:tsw}
\ee
in the SM with three families. 
From now on, $B(t \to \dots)$ for a generic decay channel,
stands for  the quantity:
\be
B(t \to \dots) = \frac{\Gamma(t \to \dots)}{\Gt}.
\ee

The two-body tree-level decay channels are the only ones that LHC could be able 
to detect in the framework of the SM.
The next less-rare processes indeed turn out to have 
rates not larger than $10^{-6}$.
\begin{figure*}[t]
\vspace*{-4cm}
\centerline{\psfig{figure=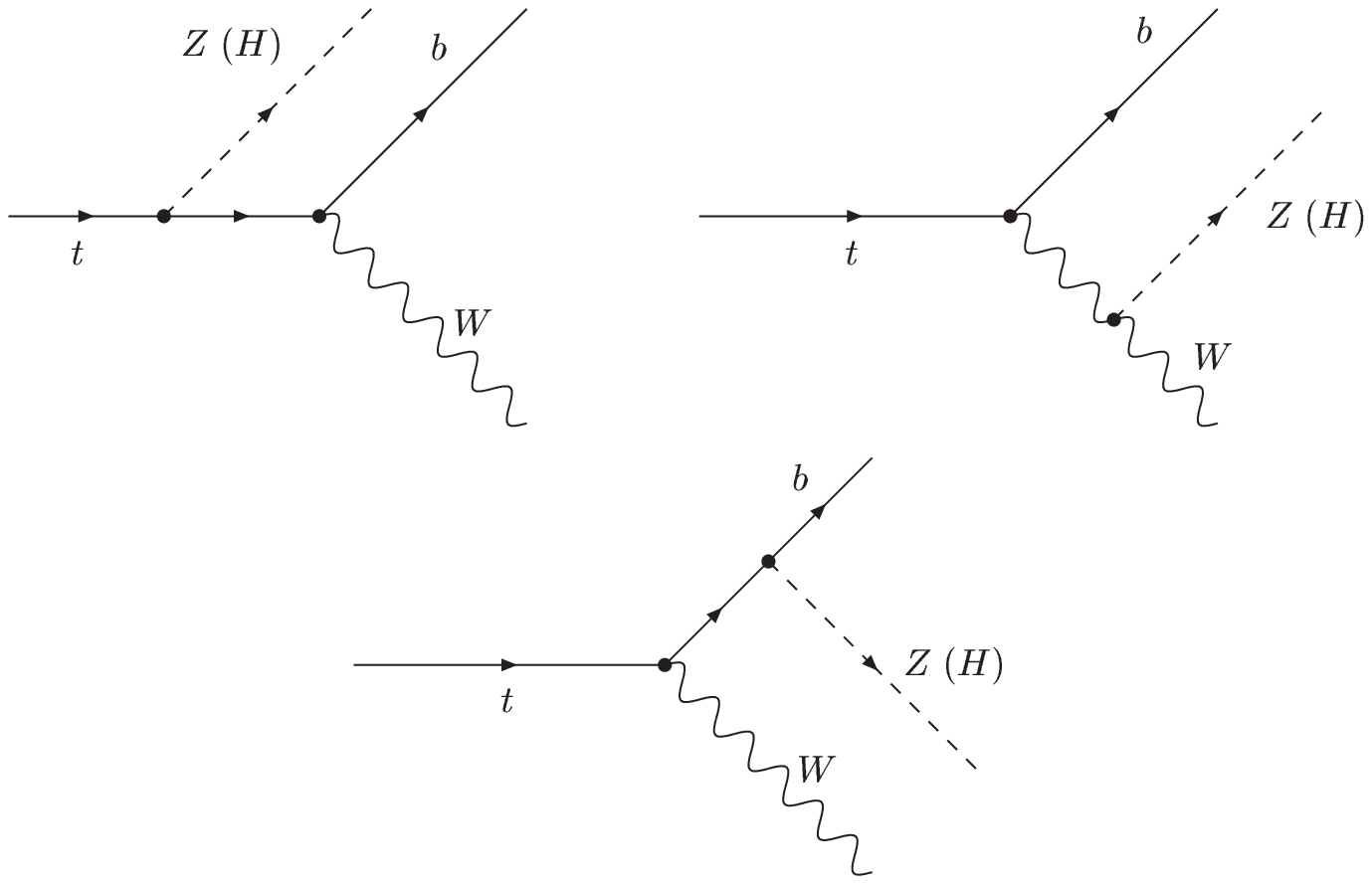,width=16cm}}
\vspace*{-12cm}
\caption{  Feynman graphs for the decay \tbwzt (\tbwhm).
 }
\label{bmf:fig1}
\end{figure*}
At tree level, the decay \tbwzt (fig.~\ref{bmf:fig1}) 
has some peculiar features,
since the process occurs near the kinematical threshold 
($\mt \sim \mw +\mz +\mb$). This fact makes crucial the $W$ and $Z$ 
finite-width effects in the theoretical prediction of the corresponding width.
In \cite{Mahlon:1995us} (see also \cite{Decker:1993wz}),
 an estimate of $B(\tbwz)$ was given on the basis
of the definition:
\be
\Gamma(\tbwz)\equiv \frac{\Gamma(t \to b \mu \nu_{\mu} ee)}
                   {B(W \to \mu \nu_{\mu}) B(Z \to ee)}
\label{bme:eq_mahlon}
\ee
involving experimentally well-observable decays.
This definition involves 14 diagrams  contributing to
the $b \mu \nu_{\mu} ee$ final state in the SM.  The estimate
for the corresponding branching ratio is:
\be
B_{cut}(\tbwz)\simeq 6 \times 10^{-7},
\label{bme:eq_m}
\ee
for $\mt=175$~GeV,
assuming a minimum cut of $0.8\mz$ on the $ee$-pair invariant mass.
This cut tries to cope with the contribution of background graphs where
the $ee$ pair rises not from a $Z$ boson but from a photon.
Recently, in \cite{bm:conti}, it has been argued that 
$B(\tbwz)$ should be defined  by including in the definition in 
eq.~(\ref{bme:eq_mahlon})
only the three (gauge-invariant) diagrams that correspond to
the process where the final state $b \mu \nu_{\mu} ee$ is truly 
mediated by a $W$ and a $Z$. Then one obtains, for $\mt=175$~GeV:
\be
B(\tbwz)=B_{res}(\tbwz)= 2 \times 10 ^{-6}.
\label{bme:res}
\ee
Such an increase in $B(\tbwz)$ is partly due to the negative interference
effects with the graphs where a photon replaces the $Z$ boson (which are
present in the previous estimate of the quantity), and partly
to the absence of any kinematical cut (which for the resonant graphs is not
needed). 

\noindent
One could also find less ambiguous definitions for $B(\tbwz)$
than the one involved in eq.~(\ref{bme:eq_mahlon}).
For instance, the new definition: 
\be
\tilde\Gamma(\tbwz)\equiv \frac{\Gamma(t \to b \mu \nu_{\mu} \nu_e \nu_e)}
                   {B(W \to \mu \nu_{\mu}) B(Z \to \nu_e \nu_e)}
\label{bme:eq_mahlon_nu}
\ee
would not require any kinematical cut, would involve a negligible
background,  and would give for $B(\tbwz)\;$ the same result as in 
eq.(\ref{bme:res}). Of course, the signature $b \mu \nu_{\mu} \nu_e \nu_e$
would not be practical from an experimental point of view, also because of the 
tiny rates involved in this decay channel. 

\noindent
Then, the authors in \cite{bm:conti} agree that an experimental
measurement of the decay rate of \tbwzt should go through the definition 
in eq.~(\ref{bme:eq_mahlon}), where the final state $b \mu \nu_{\mu} ee$
includes all the  possible backgrounds. They confirm the result
in eq.~(\ref{bme:eq_m}) as an experimental {\it effective} quantity.
\begin{figure*}[th]
\vspace*{-4.cm}
\centerline{\psfig{figure=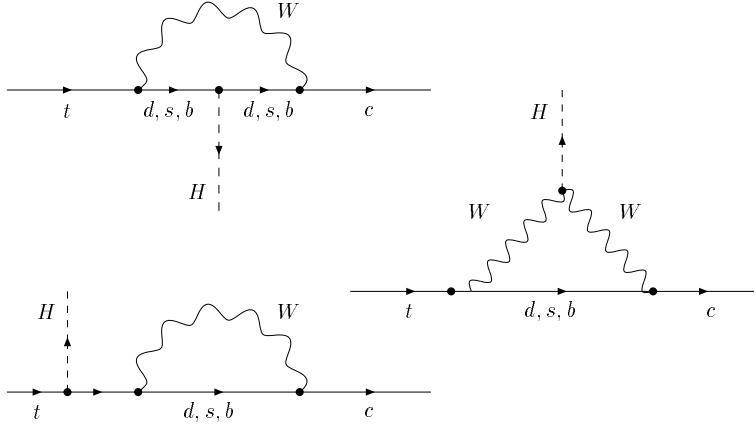,width=16cm}}
\vspace*{-11.5cm}
\caption{ Feynman graphs for the decay \tcht in the unitary gauge
($\mc=0$ is assumed).
 }
\label{bmf:fig2}
\end{figure*}
		   
If the Higgs boson is light enough, one could also have the decay
\tbwht (fig.~\ref{bmf:fig1}), although the present limits on $\mh$
strongly suppress its rate. 
For $\mh \gta 100$~GeV, one gets \cite{Mahlon:1995us} (see also 
\cite{Decker:1991cz}):
\be
B(\tbwh) \lta 7 \times 10 ^{-8}.
\ee
Finally, the decay \tcwwt is very much suppressed by a GIM factor 
$\frac{\mbb}{\mww}$ in the amplitude.
One than gets \cite{Jenkins:1997zd}:
\be
B(\tcww) \sim  10 ^{-13}.
\label{bme:tcww}
\ee
One can also consider the radiative three-body decays
\tbwgt and \tbwfm. These
channels suffer from infrared divergences, and the evaluation of their rate
requires a full detector simulation, including for instance the effects
of the detector resolution and the jet isolation algorithm.
In an idealized situation where the rate is computed in the $t$ rest frame
with a minimum cut of 10 GeV on the gluon or photon energies, 
one finds \cite{bm:mahl} (see also \cite{Decker:1993wz}):
\ba
B(\tbwg) &\simeq & 0.3,      \\
B(\tbwf) &\simeq & 3.5 \times 10 ^{-3}.
\ea

The FCNC decays \tcgm, \tcft and \tczt occur at one loop,
and are also GIM suppressed by a factor $\frac{\mbb}{\mww}$ in the amplitude.
Hence, the corresponding rates are very small \cite{bm:hewett}:
\ba
B(\tcg) &\simeq & 5 \times 10^{-11}  \\
B(\tcf) &\simeq & 5 \times 10^{-13}  \\
B(\tcz) &\simeq & 1.3 \times 10^{-13}
\label{bme:fcnc}
\ea
For a light Higgs boson, one can consider also the FCNC decay \tcht
(fig.~\ref{bmf:fig2}).
A previous evaluation of its rates \cite{bm:hewett} has now been corrected.
For $\mh \simeq 100 \;(160)$~GeV, one gets then \cite{bm:soddu,bm:hewee}:
\be
B(\tch) \simeq 0.9 \times 10^{-13} \;\; (4 \times 10^{-15}).
\ee

We conclude this section, by presenting in table \ref{bmt:sm} 
a  summary of the expected decay rates
for the main top decay channels in the SM.
\begin{table}[h]
\begin{center}
{
\begin{tabular}{|l|c|} \hline 
 channel  & $B_{SM}$  \\ \hline  \hline
  $b W$   & $1$          \\ \hline
  $s W$    & $1.6 \cdot 10^{-3}$ \\ \hline
  $d W$    & $\sim 10^{-4}$     \\ \hline
  $b W g$  &$ 0.3$ {\small($E_{g} > 10$~GeV)} \\ \hline
  $b W \gamma$  & $ 3.5\cdot 10^{-3}$ {\small $(E_{\gamma} > 10$~GeV)} \\ \hline
 $b WZ$ & $2\cdot 10^{-6}$ \\ \hline
  $c W^+ W^-$ & $\sim 10^{-13}$  \\ \hline
  $b W^+ H$ & $< 10^{-7}$ \\ \hline
 $q g$  & $5\cdot 10^{-11}$ \\ \hline
 $q \gamma$ & $5\cdot 10^{-13}$ \\ \hline
 $q Z$ & $1.3\cdot 10^{-13}$ \\ \hline
  $c H$ &  $< 10^{-13}$ \\ \hline

\end{tabular}
}
\caption{ Branching ratios for the main SM top decay channels. }
\label{bmt:sm}
\end{center}
\end{table}

\section{Beyond the Standard Model (BSM) decays}

The fact that a measurement of the top width is not available and
the branching ratio $B(\tbw)$ is a model dependent quantity  makes the present
experimental constraints on the top BSM decays  quite weak.
Hence, the large value of $\mt$ allows to consider the possibility 
of $t$ decays into new massive states with branching fraction of 
order $B(\tbw)$.
Apart from the production of new final states with large branching fractions, 
we will see that  new physics
could also give rise to a considerable increase in the rates of 
many decay channels that in the SM framework are under the threshold of 
observability. 

\subsection{$4^{th}$ fermion family}
Extending the SM  with a $4^{th}$ fermion family can alter considerably a few
$t$ decay channels. First of all, when adding a $4^{th}$ 
family to the CKM matrix
the present constraints on the $|V_{tq}|$
elements are considerably relaxed.  In particular,
$|V_{ts}|$ and   $|V_{td}|$ can grow up
to about 0.5 and 0.1, respectively~\cite{Caso:1998tx}.
Correspondingly, assuming $|V_{tb}|\sim 1$ for the sake of normalization, 
one can have up to :
\be
B_4(\tsw) \sim 0.25 \; \; {\rm and} \; \; B_4(\tdw) \sim 0.01,
\ee
to be confronted with the SM expectations in eq.~(\ref{bme:tsw}).

The presence of a $4^{th}$ fermion family could also show up in the 
$t$ direct decay into a heavy $b'$ quark with a relatively small 
mass ($m_{b'} \sim 100$~GeV) \cite{Atwood:1997iz}. This decay would 
contribute to the \tcwwt channel.
The corresponding rate would be:
\begin{eqnarray}
 {\rm B}(t \to W^+ b'(\to W^- c)) \sim 10^{-3}\;(10^{-7}) 
\quad {\rm at} \quad m_{b'} = 100 \;(300) \;\;{\rm GeV},
\label{bme:tcwwq}
\end{eqnarray}
to be confronted with the SM prediction in eq.~(\ref{bme:tcww}).
\subsection{Two Higgs Doublet  models (2HDM's)}
The possibility that the electroweak symmetry breaking involves more
than one Higgs doublet is well motivated theoretically.
In particular, three classes of two Higgs doublet models have 
been examined in connection with rare top decays, called model I, II
and III. The first two are characterized by an {\it ad hoc}
discrete symmetry which forbid tree-level FCNC~\cite{bm:glashow}, 
that are strongly 
constrained in the lightest quarks sector. In model I  and model II,
the up-type quarks and down-type quarks couple to the same scalar 
doublet and to two different doublets, respectively 
(the Higgs sector of the  MSSM is an example of model II). 
In model III \cite{bm:hall,bm:hou}, the above discrete symmetry 
is dropped and tree-level FCNC are allowed. In particular,
a tree-level coupling $tcH$ is predicted with a coupling constant 
$\lta \frac{\sqrt{\mt\mc}}{v}$ (where $v$ is the Higgs vacuum 
expectation value).

\noindent
Since enlarging the Higgs sector automatically implies the presence of 
charged Higgs bosons in the spectrum, 
one major prediction of these new frameworks is  the decay $t\to bH^+$, 
with possibly competitive rates with $B(\tbw)$ for $m_{H^+}\lta 170$ GeV.
In the MSSM, one has $B(t\to bH^+)\sim 1$, both at small and large values
of $\tan\beta$. 
 If $m_H < m_t - m_b$, one  expects
$H^+ \to \tau^+ \nu$ (favored for large $\tan\beta$) and/or
$H^+ \to c \bar{s}$ (favored for small $\tan\beta$) to be the 
dominant decays.
Hence, for $\tan\beta >1$ and 
 $H^+\to \tau^+ \nu$  dominant,  one  can look  for the channel
$t\to bH^+$ by studying a possible excess in the $\tau$ lepton 
signature from the $t$ pair production \cite{bm:cdf-tau}.
On the other hand,
if $\tan\beta < 2$ and $m_H > 130$~GeV, the
large mass (or coupling) of the $t$-quark causes
${\rm B}(H^+ \to t^{*} \bar b \to W^+ b \bar b)$ to exceed
${\rm B}(H^+ \to c \bar s)$ (see~\cite{Ma:1998up} 
for details).
As a consequence, new interesting signatures at LHC like 
leptons plus multi-jet channels  with four $b$-tags, coming from
the gluon-gluon fusion process $gg\to t\bar b H^-$,
followed by the $H^-\to \bar t b$ decay, 
have been studied \cite{bm:stirling}. These processes could 
provide a viable signature over a limited but interesting range of the
parameter space.  

\noindent
One should recall however that both $B(t\to bH^+)$ 
and  $B(H^+\to W^+b\bar{b})$ are very sensitive
to higher-order corrections, which are highly model dependent
\cite{bm:coarasa,bm:junbi}.

In model III, the tree-level FCNC decay $t \to ch$ can occur
with branching ratios up to 10$^{-2}$ \cite{bm:hou}.
In \cite{Bar-Shalom:1998sj}, the rate for the channel 
$t \to ch \to cWW (cZZ, c\gamma\gamma)$
has been studied. Accordingly,
$B(\tcww)$ can be enhanced by several orders of magnitude
with respect to its SM value. In particular, for an on-shell decay
with $2\mw\lta m_h \lta \mt$, one can have up to $B(\tcww)\sim 10^{-4}$
from this source.  The same process was considered in a wider range of
models, where the decay \tcwwt can occur not only through 
a scalar exchange but also through a fermion or vector exchange
\cite{Atwood:1997iz}. In this framework, the fermion exchange too could 
lead to detectable rates for \tcwwm, as in eq.~(\ref{bme:tcwwq}).

In 2HDM's, the prediction for the FCNC decays
\tcgm, \tcft and \tczt can also be altered. While in models I and II
the corresponding branching fractions can not anyway approach the detectability
threshold \cite{bm:hewett}, in model III values up to
$B(\tcg) \simeq  10^{-5}$, $B(\tcf) \simeq  10^{-7}$ and 
$B(\tcz) \simeq  10^{-6}$  are predicted \cite{bm:reina}.

By further extending the 2HDM's sector and including Higgs triplets,
one can give rise to a vertex $HWZ$ at tree level in a consistent way
\cite{bm:diazcruz}. Accordingly, the \tbwzt decay can be mediated
by a charged Higgs (coupled with $\mt$) that can enhance 
the corresponding branching fraction
up to $B(\tbwz) \sim 10 ^{-2}$. Large enhancements can also be expected
in similar models for the channels $t \to sWZ\;$and $t \to dWZ$.

\subsection{Minimal Supersymmetric Standard Model (MSSM)}
Supersymmetry could affect the $t$ decays in different ways
[here, we assume the MSSM framework \cite{bm:mssm}, with
(or without, when specified) $R$ parity conservation].

First of all, two-body decays into squarks and gauginos, such as
$t \to \tilde t_1 \tilde g$, $t \to \tilde b_1 \tilde \chi_1^+$,
$t \to \tilde t_1 \tilde \chi_1^0$, could have branching ratios
of order $B(\tbw)$, if allowed by the phase space
(see, i.e., \cite{bm:guasch-three} for references).
QCD corrections to the channel $t \to \tilde t_1 \tilde g$
have been computed in \cite{bm:zhu} and found to increase 
its width up to values even larger than $\Gamma(\tbw)$.
Three-body $t$ decays in supersymmetric particles were
surveyed in \cite{bm:guasch-three}.

The presence of light top and bottom squarks, charginos and 
neutralinos in the MSSM spectrum could also give rise to a $CP$
asymmetry of the order $10^{-3}$ in the partial widths 
for the decays $t \to b W^+\;$ and $\bar t \to \bar b W^-$
\cite{bm:aoki}.

Explicit $R$-parity violating interactions \cite{bm:roy} could 
provide new flavor-changing $t$ decays, both at tree-level
(as in the channels $t \to \tilde \tau b$ 
and $t \to \tau b \tilde \chi_1^0\;$ \cite{bm:magro})
and at one loop  (as in $t \to c \tilde \nu\;$ \cite{bm:soni}),
with observable rates. For instance, $B(t \to c \tilde \nu) \sim
10^{-4}-10^{-3}\;$ in particularly favorable cases.

Another sector where supersymmetric particles could produce
crucial changes concerns  the one-loop FCNC decays
\tcgm, \tcfm, \tczt and \tcht, which in the SM are unobservally
small. In the MSSM with universal soft breaking the situation is 
not much affected,
while, by relaxing the universality with a large flavor mixing between the
2$^{nd}$ and 3$^{rd}$ family only, one can reach values such as
\cite{bm:petronzio,bm:lopez}:
\ba
B_{MSSM}(\tcg) &\sim &  10^{-6}  \\
B_{MSSM}(\tcf) &\sim &  10^{-8}  \\
B_{MSSM}(\tcz) &\sim &  10^{-8},
\label{bme:fcncsusy}
\ea
which anyhow are still not observable.
The introduction of $B$-violating couplings in broken $R$-parity
models could on the other hand give large enhancements \cite{bm:yang},
and make some of these channels observable.
The corresponding upper limits on branching ratios get then:
\ba
B_{R\!\!\!\! /}(\tcg) &\sim &  10^{-3}  \\
B_{R\!\!\!\! /}(\tcf) &\sim &  10^{-5}  \\
B_{R\!\!\!\! /}(\tcz) &\sim &  10^{-4}.
\label{bme:fcncbr}
\ea

A particularly promising channel is the FCNC decay $t \to c h$ in 
the framework of MSSM, where $h = h^0,H^0,A^0$ is any of the 
supersymmetric neutral Higgs bosons \cite{bm:sola}. 
By including the 
leading MSSM contributions to these decays (including gluino-mediated FCNC
couplings),
one  could approach the detectability
threshold, especially in the case of the light CP-even Higgs boson, 
for which one can get up to:
\be
B_{MSSM}(t \to c h^0) \sim 10^{-4}.
\ee

\section{LHC potential for  top rare decays}
An extensive analysis of the LHC  potential
for detecting top rare decays has been performed by both the LHC experiments
ATLAS and CMS \cite{bm:lhc-report}.
In the framework of the SM, the top rare decays (that is any channel
different from $t\to q W$) turn out to be definitely
below the threshold for an experimental analysis at LHC.
On the other hand, LHC experiments will be able to probe quite a few 
predictions of possible extensions of the SM.

An extended Higgs sector will be looked for through
the tree-level decay $t \to b H^+$. ATLAS estimates its sensitivity
to  this channel in the MSSM, 
 through an excess in the tau lepton signal,
to be around  B$(t \to H^{+}b)$ = 3\% (that is almost 4 times better than 
what expected from Run 2 at the Tevatron).
This would allow to probe
all values of $m_{H^{\pm}}$ below $m_{t} - 20$ GeV over most 
of the $\tan \beta$ range.
For low $\tan \beta$, the complementary decay mode 
$H^{\pm} \to cs$ has been considered.
In the mass range 110 $< H^{\pm} <$
130 GeV, the $H^{\pm}$ peak can be reconstructed
and separated from the dominant $W \to jj$ 
background.

\noindent
For CMS, using the tau excess signature, 
the expected 5$\sigma$ discovery range for 10 fb$^{-1}$ in 
the MSSM ($m_A, \tan\beta$) parameter space is $m_A <$ 110 GeV,
 for all $\tan\beta$ values, and
 somewhat extended ($m_A \lta$ 140), for $\tan\beta \lta$ 2.  
 
\noindent
Other interesting signatures like  
$H^{\pm} \to hW^{\ast}$, $H^{\pm} \to AW^{\ast}$ and 
$H^{\pm} \to bt^{\ast} \to bbW$ 
are very promising in particular parameter ranges, but have not yet been
thoroughly investigated.

ATLAS has studied its sensitivity to
the radiative decay $t \to WbZ$. This has been found to be at most of 
the order
10$^{-4}$, that is insufficient for the study of a SM signal ($\sim 10^{-6}$), 
but possibly useful for exploring the predictions of some extended Higgs-sector
model, for which B($t \to WqZ)\lta 10^{-2}$.
On the other hand, the radiative Higgs decay  $t \to WbH$
seems out of the reach of LHC in any realistic model.

The LHC reach for the FCNC decays 
$t \to qZ$, $t \to q\gamma \; $ and $t \to qg\; $ has also been thoroughly
investigated.
Apart from  the $t \to qg\; $, which is completely overwhelmed by the hadronic
background, both ATLAS and CMS have a sensitivity of about $2 \times 10^{-4}$
to the $t \to qZ$ channel, while the CMS reach for the $t \to q\gamma \; $
channel is about $3.4 \times 10^{-5}$, that is slightly better than the ATLAS
sensitivity ($1.0 \times 10^{-4}$),
assuming an integrated 
luminosity of 100 fb$^{-1}$. These thresholds are too high
to test the predictions of the models reviewed here. On the other hand,
they could be largely
sufficient to detect some manifestation of possible FCNC anomalous couplings
in the top sector \cite{bm:lhc-report}.
\newpage



\begin{thebibliography}{99}

\bibitem{Caso:1998tx}
C. Caso et al, Eur.\ Phys.\ J.\ {\bf C3}, 1 (1998), 
  and 1999 off-year partial update for the 2000 edition available on 
  the PDG WWW pages (URL: http://pdg.lbl.gov/). 
\vspace{-2.5mm}

\bibitem{Mahlon:1995us}
G.~Mahlon and S.~Parke,
Phys.\ Lett.\  {\bf B347}, 394 (1995)
[hep-ph/9412250].
\vspace{-2.5mm}

\bibitem{Decker:1993wz}
R.~Decker, M.~Nowakowski and A.~Pilaftsis,
Z.\ Phys.\ C {\bf 57}, 339 (1993)
[hep-ph/9301283].
\vspace{-2.5mm}

\bibitem{bm:conti}
G.~Altarelli, L.~Conti and V.~Lubicz,
Phys.\ Lett.\ B {\bf 502}, 125 (2001)
[hep-ph/0010090].
\vspace{-2.5mm}

\bibitem{Decker:1991cz}
R.~Decker, M.~Nowakowski and A.~Pilaftsis,
Mod.\ Phys.\ Lett.\ A {\bf 6}, 3491 (1991)
[Erratum-ibid.\ A {\bf 7}, 819 (1991)].
\vspace{-2.5mm}

\bibitem{Jenkins:1997zd}
E.~Jenkins,
Phys.\ Rev.\  {\bf D56}, 458 (1997)
[hep-ph/9612211].
\vspace{-2.5mm}

\bibitem{bm:mahl}
G.~Mahlon,
[hep-ph/9810485].
\vspace{-2.5mm}

\bibitem{bm:hewett}
G.~Eilam, J.L.~Hewett and A.~Soni, 
Phys.\ Rev.\ {\bf D44}, 1473 (1991).
\vspace{-2.5mm}

\bibitem{bm:soddu}
B.~Mele, S.~Petrarca, A.~Soddu, 
Phys.\ Lett.\ {\bf B435}, 401 (1998), [hep-ph/9805498].
\vspace{-2.5mm}

\bibitem{bm:hewee}
G.~Eilam, J.L.~Hewett and A.~Soni, 
Erratum-Phys.\ Rev.\ {\bf D59}, 039901 (1999).
\vspace{-2.5mm}

\bibitem{Atwood:1997iz}
D.~Atwood and M.~Sher,
Phys.\ Lett.\  {\bf B411}, 306 (1997)
[hep-ph/9707229].
\vspace{-2.5mm}

\bibitem{bm:glashow}
S.~Glashow and S.~Weinberg,
Phys.\ Rev.\ {\bf D15}, 1958 (1977).
\vspace{-2.5mm}

\bibitem{bm:hall}
A.~Antaramian, L.~J.~Hall and A.~Rasin,
Phys.\ Rev.\ Lett.\  {\bf 69}, 1871 (1992)
[hep-ph/9206205]; \\
L.J.~Hall and S.~Weinberg,
Phys.\ Rev.\ {\bf D48}, R979 (1993), [hep-ph/9303241]; \\
M.~Luke and M.J.~Savage,
Phys.\ Lett.\ {\bf B307}, 387 (1993), [hep-ph/9303249].
\vspace{-2.5mm}

\bibitem{bm:hou}
W.S.~Hou, 
Phys.\ Lett.\ {\bf B296}, 179 (1992).
\vspace{-2.5mm}


\bibitem{bm:cdf-tau}
T.~Affolder {\it et al.}  [CDF Collaboration],
Phys.\ Rev.\ D {\bf 62}, 012004 (2000)
[hep-ex/9912013].
\vspace{-2.5mm}

\bibitem{Ma:1998up}
E.~Ma, D.~P.~Roy and J.~Wudka,
Phys.\ Rev.\ Lett.\  {\bf 80}, 1162 (1998)
[hep-ph/9710447].
\vspace{-2.5mm}

\bibitem{bm:stirling}
D.~J.~Miller, S.~Moretti, D.~P.~Roy and W.~J.~Stirling,
Phys.\ Rev.\ D {\bf 61}, 055011 (2000)
[hep-ph/9906230].
\vspace{-2.5mm}

\bibitem{bm:coarasa}
J.A.~Coarasa, J.~Guasch and J.~Sola, hep-ph/9903212; \\
J.A.~Coarasa, J.~Guasch, W.~Hollik and J.~Sola,
Phys.\ Lett.\ {\bf B442}, 326 (1998), [hep-ph/9808278].
\vspace{-2.5mm}

\bibitem{bm:junbi}
Xiao-Jun Bi, Yuan-Ben Dai, Xiao-Yuan Qi,
Phys.\ Rev.\ {\bf D61}, 015002 (2000),
[hep-ph/9907326].
\vspace{-2.5mm}

\bibitem{Bar-Shalom:1998sj}
S.~Bar-Shalom, G.~Eilam, A.~Soni and J.~Wudka,
Phys.\ Rev.\  {\bf D57}, 2957 (1998)
[hep-ph/9708358]; \\
S.~Bar-Shalom, G.~Eilam, A.~Soni and J.~Wudka,
Phys.\ Rev.\ Lett.\  {\bf 79}, 1217 (1997), [hep-ph/9703221]; \\
J.~L.~Diaz-Cruz, M.~A.~Perez, G.~Tavares-Velasco and J.~J.~Toscano,
Phys.\ Rev.\ D {\bf 60}, 115014 (1999)
[hep-ph/9903299].
\vspace{-2.5mm}

\bibitem{bm:reina}
D.~Atwood, L.~Reina and A.~Soni,
Phys.\ Rev.\ {\bf D55}, 3156 (1997), [hep-ph/9609279];\\
see also
L.~Reina, talk given at the 
Fermilab Thinkshop on Top Physics at Run II (Oct 19-21 1998).
\vspace{-2.5mm}

\bibitem{bm:diazcruz}
J.~L.~Diaz Cruz and D.~A.~Lopez Falcon,
Phys.\ Rev.\ D {\bf 61}, 051701 (2000)
[hep-ph/9911407].
\vspace{-2.5mm}

\bibitem{bm:mssm}
For a review, see for instance, H.E.~Haber and G.L.~Kane,
Phys.\ Rep.\ {\bf 117}, 75 (1985).
\vspace{-2.5mm}

\bibitem{bm:guasch-three}
F.~M.~Borzumati and N.~Polonsky,
hep-ph/9602433; \\
J.~Guasch and J.~Sola,
Z.\ Phys.\  {\bf C74}, 337 (1997)
[hep-ph/9603441].
\vspace{-2.5mm}

\bibitem{bm:zhu}
S.H.~Zhu and L.Y.~Shan, hep-ph/9811430.

\vspace{-2.5mm}

\bibitem{bm:aoki}
S.~Bar-Shalom, G.~Eilam, A.~Soni, J.~Wudka, 
Phys.\ Rev.\ {\bf D57}, 1495 (1998) [hep-ph/9708357];
M.~Aoki, N.~Oshimo,
Mod.\ Phys.\ Lett.\ {\bf A13}, 3225 (1998) [hep-ph/9808217]. 
\vspace{-2.5mm}

\bibitem{bm:roy}
For  reviews on $R\!\!\! /$, see e.g. : 
D.P.~Roy, Pranama {\bf 41}, S333 (1993) [hep-ph/9303324]; \\
G.~Bhattacharyya, Nucl.\ Phys.\  Proc.\ Suppl.\ {\bf 52A}, 83 (1997).
\vspace{-2.5mm}

\bibitem{bm:magro}
T.~Han and M.~B.~Magro,
Phys.\ Lett.\ B {\bf 476}, 79 (2000)
[hep-ph/9911442].
\vspace{-2.5mm}

\bibitem{bm:soni}
S.~Bar-Shalom, G.~Eilam and A.~Soni,
Phys.\ Rev.\ {\bf D60}, 035007 (1999) [hep-ph/9812518].
\vspace{-2.5mm}

\bibitem{bm:petronzio}
G.M.~de Divitiis, R.~Petronzio, L.~Silvestrini,
Nucl.\ Phys.\ {\bf B504}, 45 (1997), [hep-ph/9704244].
\vspace{-2.5mm}

\bibitem{bm:lopez}
J.L.~Lopez, D.V.~Nanopoulos and R.~Rangarajan,
Phys.\ Rev.\ {\bf  D56}, 3100 (1997), [hep-ph/9702350].
\vspace{-2.5mm}

\bibitem{bm:yang}
J.M.~Yang, B.-L.~Young and X.~Zhang,
Phys.\ Rev.\ {\bf  D58}, 055001 (1998).
[hep-ph/9705341],
\vspace{-2.5mm}

\bibitem{bm:sola}
J.~Guasch and J.~Sola,
Nucl.\ Phys.\ B {\bf 562}, 3 (1999)
[hep-ph/9906268].
\vspace{-2.5mm}

\bibitem{bm:lhc-report}
M.~Beneke, I.~Efthymiopoulos, M.L.~Mangano, J.~Womersley (conveners), et al. 
(The Top Physics Working Group), {\it Top Quark Physics}, 
hep-ph/0003033, \\
to appear in the Report of the 
``1999 CERN Workshop on SM physics (and more) at the LHC",
CERN-TH/2000-100.

\vspace{-2.5mm}

\end{thebibliography}
\end{document}